\input harvmac
\input epsf
\Title{\vbox{\hbox{HUTP-99/A008}\hbox{hep-th/99xxxxx}}}
{\vbox{\hbox{\centerline{$PSL(n|n)$ Sigma Model as a Conformal Field
Theory}}}}
\centerline{Michael Bershadsky$^{(a)}$, Slava Zhukov$^{(a)}$ 
and Arkady Vaintrob$^{(b)}$}
\vskip .2in
\centerline{$^{(a)}$\it Lyman Laboratory of Physics, Harvard
University, Cambridge, MA 02138}
\vskip .2in
\centerline{$^{(b)}$\it Department of Mathematical Sciences, 
New Mexico State University,
Las Cruces, NM 88003}
\vskip .5in

\centerline{\bf Abstract}

We discuss the sigma model on the $PSL(n|n)$ supergroup manifold. We
demonstrate that this theory is exactly conformal. The chiral algebra of
this model is given by some extension of the Virasoro algebra, similar to
the $W$ algebra of Zamolodchikov. We also show that all group invariant
correlation functions are coupling constant independent and can be computed
in the free theory. The non invariant correlation functions are highly
nontrivial and coupling dependent. At the end we compare two and
three-point correlation functions of the $PSL(1,1|2)$ sigma model with the
correlation functions in the boundary theory of $AdS_3 \times S^3$ and find
a qualitative agreement.

\vskip .2in
\Date{February, 1999}

\newsec{Introduction}

In this paper we discuss two-dimensional nonlinear
sigma models on the supergroup manifolds $PSL(n|n)$.
The interest in these models is motivated by the  
recent discoveries in the string theory. 
It was long suspected by Polyakov and others (see, for example
\ref\pol{A. Polyakov,
``The wall of the cave'', hep-th/9809057} and references therein) 
that  gauge theory can be described by some version of
string theory. The first concrete example of this was recently 
suggested in \ref\mald{J. Maldacena, ``The Large N Limit of Superconformal
Field Theories and Supergravity'', hep-th/9711200}. The strings propagate
in the 
$AdS$-type supergravity background. 
The ${\cal N}=4$ supersymmetric gauge theory ``lives'' on the 
boundary of the  $AdS$ space and provides boundary 
conditions for the bulk gravity 
theory. The supergravity describes the large coupling limit
of the gauge theory (literally speaking $g^2 N= \infty$ limit).
The $1/N$ corrections can be identified with string loops.   
The string theory appears to be {\it critical} with constant dilaton which
implies that the boundary theory is conformal.

\nref\kut{A. Giveon, D. Kutasov and N. Seiberg,
{\it Comments on String Theory on $AdS_3$}, hep-th/9806194}

\nref\rah{J. Rahmfeld, A. Rajaraman, {\it The GS String Action on 
$AdS_3 \times S^3$ with Ramond-Ramond Charge}, hep-th/9809164}
\nref\pess{Igor Pesando, {\it The GS Type IIB Superstring Action on 
$AdS_3 \times S^3 \times  T^4$}, hep-th/9809145}
\nref\jpar{J. Park and S.-J. Rey, 
{\it Green-Schwarz Superstring on $AdS_3 \times S^3$}, hep-th/9812062}

\nref\ts{R. R. Metsaev and A.A. Tseytlin, Nucl. Phys. {\bf B533} 1998 109, 
hep-th/9805028}

\nref\kall{R. Kallosh, J. Rahmfeld and A. Rajaraman, {\it Near Horizon
Superspace},hep-th/9805217}
\nref\kallt{R. Kallosh, J. Rahmfeld, Phys. Lett. {\bf B443} 143,
hep-th/980838}
\nref\pes{Igor Pesando, 
{\it A $k$ Gauge Fixed Type IIB Superstring Action on $AdS_{5} \times
 S_{5}$}, hep-th/9808020}
\nref\roz{A. Rajaraman and M. Rozali, 
{\it On the Quantization of the GS String on $AdS_5 \times S^5$},
hep-th/9902046}

The immediate problem is to understand string theory in the $AdS$-type
backgrounds. Unfortunately, the NSR formalism does not seem to be suitable
for the description of RR backgrounds. The appearance of the RR vertex
operators introduces an arbitrary number of cuts ruining the NSR worldsheet.
One can define/compute the scattering amplitudes of several RR vertex
operators but it is unclear how to describe a condensate or a background of
RR fields.  The GS formalism seems to be more appropriate for that.  This
reminds us of the sigma model with the target space being the
supermanifold.  Therefore, one can suggest that the $PSU(1,1|2)$, or more
generally the $PSL(n|n)=SL(n|n)/U(1)$ sigma models would naturally
appear in this description. Indeed, the appearance of the worldsheet
scalars/target space fermions is the genuine feature of the GS-type
actions. On the other hand the second order fermionic kinetic term
indicates the possible relation to RR flux.  In fact, there are suggestions
for a GS formulation of string theory based on the $PSL(n|n)$ cosets 
\refs{\rah{--}\roz}. 
Unfortunately, it is unclear how to quantize this string theory.
The above provides more than enough motivation to study these sigma models.

To be a bit more precise, one can say that $PSU(1,1|2)$ sigma model
is related to string theory propagating in $AdS_3 \times S^3 \times M^4$
background. The NSR formulation of string theory propagating on 
this background was given in \kut. 
The classical formulation of GS superstring was discussed in
\refs{\rah {--}\jpar}. 
The $PSU(1,1|2)$ sigma model is not a string theory yet
but it is going to be an important ingredient in the construction.
Such string theory was recently proposed in 
\ref\bvw{N. Berkovits, C. Vafa, E. Witten,
{\it Conformal Theory of AdS Background with Ramond-Ramond Flux}, 
hep-th/9902098}.

Similarly, the sigma model on a $PSU(2,2|4)$ is related to the string theory
on the $AdS_5 \times S^5$ background.  The latter can be regarded as the
bosonic part (body) of a
{\it quotient} superspace of a supergroup by the subgroup 
\eqn\quo{{\cal T}={PSU(2,2|4) \over SO(4,1) \times SO(5)}}
The $SO(4,1) \times SO(5)$ action is an isometry without fixed
points.  Notice, that there is no GKO construction and therefore it is
not obvious that the quotient \quo\ leads to the conformal theory. 
However the subgroup $SO(4,1) \times SO(5)$ turns out to be very special
and one can show that the sigma model
on a quotient space is one loop finite.
Moreover, we expect that sigma model on the quotient space is
exactly conformal in all oders in perturbation theory.
The classical formulation of GS string theory based on this quotient was
discussed in \ts, \refs{\kall{--}\roz}.  We believe that understanding the
$PSU(2,2|4)$ sigma model may shed some light on how the string theory on
$AdS_5 \times S^5$ background can be quantized.

It is clear, even without making any calculations, that
the $PSL(n|n)$ sigma models are quite remarkable. The dual 
Coxeter number $C_V$ vanishes for these groups and therefore
the one loop beta function, which is 
proportional to $C_V$, is also zero.
Hence, one may suspect that these theories are exactly conformal.
Moreover, the supermanifold $PSL(n|n)$ is in a sense a Calabi-Yau
supermanifold (see \ref\st{S. Sethi,
Nucl.Phys. {\bf B430} (1994) 31-50}). Namely,
the Ricci tensor is identically equal to zero $R_{ij} \sim C_V
g_{ij}$. 
The Ricci flatness implies that the theory is one loop finite.
In \st\ the finiteness of the theory was ensured by ${\cal N}=2$
supersymmetry, in our case the theory is
conformal {\it without} worldsheet supersymmetry.  
The idea of thinking about $PSL(n|n)$ as
a Calabi-Yau manifold could be very fruitful for possible future
applications. For example, one may try to construct new examples of
supermanifolds that lead to exactly conformal field theories
by tensoring several $PSL(n|n)$ models and then taking a quotient by  
certain subgroups. Very little is known about sigma models on 
supermanifolds and potential applications are enormous.

In this paper we study the sigma model on a supergroup $PSL(n|n)$ in
perturbation theory (large volume expansion). Clearly, there is no choice
of real structure on the group for which the metric is positive
definite. Nevertheless, we assume that the theory can be well defined. For
most part of the text our discussion is going to be purely algebraic such
that it does not require a choice of the real structure on the group. We
specify the signature at the very end when we discuss the relation to
$AdS_3 \times S^3$. We prove that the theory is exactly conformal to all
orders in perturbation theory for any values of the coupling constant.  We
also prove that the correlation functions of the group invariant
combinations of operators are given by the gaussian integration and the
interaction vertices do not contribute to these calculations. This
observation does not make the theory trivial. There are plenty of
correlation functions that have a non-trivial dependence on the sigma model
coupling.  These are the correlation functions that are not invariant under
the group action. The spectrum of conformal primary fields is classified by
the representations of the left/right multiplication symmetry group --
$V_{(\Lambda_L, \Lambda_R)}$.  We will describe a certain class of
operators (analogous to vertex operators of string theory) corresponding to
representations $(\Lambda_L, \Lambda_R)$ such that all Casimir operators
$\hat C^{(N)}$ have the same eigenvalues for both of them ($
C^{(N)}_{\Lambda_L}= C^{(N)}_{\Lambda_R}$). We conjecture that conformal
dimensions of these operators are equal to $\Delta=\bar \Delta=\lambda^2
C^{(2)}/2$, where $\lambda$ is the sigma model coupling constant and
$C^{(2)}$ is the eigenvalue of quadratic Casimir.  These operators are very
similar to the momentum modes of $c=1$ system on a circle.  We also find
that $PSL(n|n)$ sigma model has a chiral algebra which is an extension of
the Virasoro algebra (similar to W-algebras of Zamolodchikov). Very little
is known about such W-algebras and it could be a very promising direction
for future studies.

In the next section we discuss the definition of $PSL(n|n)$ sigma model, as
well as some properties of $PSL(n|n)$ group. In section 3 we study our 
sigma model in perturbation theory and show that it is exactly
conformal using the background method.
Then we analyze the possible quantum corrections to the equation
$\bar \partial T_{zz}=0$ and show that they vanish. 
We also present some examples of correlation functions involving
currents. In section 4 we present some non-perturbative arguments 
based on localization.
In section 5 we describe the chiral algebra of the $PSL(n|n)$ sigma model.
We believe that the presence of this algebra should be important for
solving
the theory. Finally, in the last section we discuss physical operators and
try to make contact with the string theory on $AdS_3 \times S^3$.

\bigskip
While this paper was in a final stage of preparation we received a
paper \bvw, that partially overlaps ours.  

\newsec{Principal Chiral Field}

The $PSL(n|n)$ principal chiral field is a two-dimensional 
non-linear sigma 
model with the fields $G(x)$ taking values in a supergroup $PSL(n|n)$.
The action for the principal chiral model is  
\eqn\act{S[G]={1 \over 4 \pi \lambda ^2}\int {\rm Str} 
\big( |G^{-1} d G|^2  \big) d^2 x  ~.}
It is invariant with respect to both left and right multiplications
$G(x) \rightarrow U_L G(x) U_R ^{-1}$, 
the corresponding conserved currents are
\eqn\cur{J_R=G^{-1}dG ~,~~J_L=dG G^{-1}~.}
In its turn, the left current is invariant under the right multiplication
symmetry, 
while the right current is invariant under the left multiplication.
Under the left (right) multiplication the left (right) current transforms
by conjugation $J_{L,R} \rightarrow U_{L(R)} J_{L(R)}U_{L(R)} ^{-1}$.
The presence of this symmetry (especially its fermionic part) makes it
difficult to define a theory.

Let us remind the reader what happens in the case of 
conventional bosonic sigma 
models. As usual, correlation functions are normalized by the
partition function
\eqn\norm{\langle \prod_i {\cal O}_i  \rangle={ \int [Dg] \prod {\cal O}_i
e^{-S} 
 \over \int [Dg] e^{-S} }~.}
The integral runs over the space of maps from our worldsheet (say, the
sphere) into the group manifold. The symmetry groups $G_L$/$G_R$ act on
this space by left/right multiplications. Therefore one can try to
factorize the path integral into the integral over the space of orbits and
a finite dimensional integration along the orbit. In the case of correlation
functions invariant under either $G_L$ or $G_R$ action, integration along
the corresponding orbit introduces a multiplicative factor (the volume of
the group), which cancels in the numerator and denominator.  For 
non-invariant quantities the group integration projects on the invariant
subspace.

One way to compute correlation functions is to use perturbation theory.  To
build a perturbation theory one is forced to fix the ``false'' vacuum to
expand around. For example, one can impose the condition that at infinity the field
$G(x)$ approaches some fixed element $G_{0}$.  This choice clearly
breaks the left/right multiplication symmetry.  Again, in the case of
correlation functions invariant under either $G_L$ or $G_R$ action, the
remaining integral over $G_{0}$ can be thought of as an integral along
the orbit of a symmetry action and hence gives the same volume factor
which cancels. The quantum field around $G_{0}$ is a Goldstone boson
of this broken symmetry and therefore massless. Consequently, we run into
infrared problems. To cure those one can add a small
mass term, or the potential around $G_{0}$, or work in the finite
volume.  However, the above correlation functions are perturbatively well
defined, i.e., IR finite
\ref\eli{S. Elitzur, IAS preprint (1979)}
\ref\dav{F. David, Comm. Math. Phys. {\bf 81} 149 (1981)} 
and therefore one may trust the perturbative calculations.  Still, one has
to be careful making conclusions based on them. The perturbation theory 
does not feel the global properties of the group as it is built as an
expansion
around a particular point.

It is instructive to consider a simple example of a free scalar 
field theory. The theory is invariant with respect to a constant shift $X
\rightarrow X+c$ (zero mode).
The correlation function of the vertex operators requires an IR cutoff $m$
\eqn\ex{\langle e^{ikX(z_1)}...  e^{ikX(z_n)} \rangle=m^{(\sum k_i)^2
/2}\prod
|z_i-z_j|^{k_i k_j}~.}
This correlation function is invariant under the symmetry 
$X \rightarrow X+c$ only if $\sum k_i=0$ (otherwise it gets multiplied by a
phase).
Invariant correlation functions are independent on the IR cutoff, while
non-invariant correlation functions vanish.

Trying to repeat the above steps 
in the case of a supergroup we find that the naive integration along the 
orbit produces zero -- the volume of the supergroup. 
However we still want to define
correlation functions by normalizing them by the partition function as in
\norm.  Formally, the numerator and denominator vanish as the consequence
of
the zero group volume, but we can take a sensible limit to define the
ratio. First we need to break the left/right multiplication symmetry, say
by the same condition that at infinity the group element $G(x)$ approaches
some fixed element $G_{0}$.
The fluctuations around the ``false'' vacuum $G_{0}$ are massless bosonic
and
fermionic degrees of freedom. The latter give zero (due to fermionic zero
modes) while the massless bosons give rise to the infrared problems.
Again, adding a small mass term or the potential around $G_{0}$ fixes 
both problems
and allows one to define the correlation function as the limit of the ratio
when the mass/IR cut-off is sent to zero. 
In fact this is exactly what is computed in perturbation theory for
the conventional principal chiral model.

The $G_L$  or $G_R$ invariant correlation functions
are again IR finite and therefore this procedure provides a consistent
way of making sense out of \norm.
The situation with non-invariant quantities is far from being well 
understood. In the bosonic case the group integration effectively
projects on the group invariant subsector. In the case of a supergroup, there is 
no well defined projection operator  and one has to be very careful
(some examples of these computations will be discussed later).

\subsec{Introduction to $GL(n|n)$, $SL(n|n)$ and $PSL(n|n)$ groups.}

The groups $GL(n|n)$, $SL(n|n)$ and $PSL(n|n)$ are
closely related to each other. 
The supergroup $GL(n|n)$ consists of real $(n|n)$ supermatrices
with non-zero superdeterminant. $SL(n|n)$ is a subgroup of $GL(n|n)$ --
simply matrices with superdeterminant equal to $1$. 
It has a normal $U(1)$ subgroup -- matrices, that are multiples of the 
identity. The $PSL(n|n)$ is a factor of 
$SL(n|n)$ by that subgroup. Unfortunately, $PSL(n|n)$ {\it does not} have a 
representation in $Mat(n|n)$.  
This $U(1) \subset SL(n|n)$ that one has to factor out to get the 
$PSL(n|n)$ group appears  as
an additional gauge symmetry in the $SL(n|n)$ principal
chiral field. This suggests a way to think about $PSL(n|n)$ principal 
chiral field as a gauge invariant subsector of the $SL(n|n)$ sigma model.

We start with properties of the above groups or, rather, their Lie 
superalgebras.  The superalgebra $gl(n|n)$ has a non-degenerate metric given 
by $g_{ij}=Str(T_i T_j)$, where $T_i$ are the generators of $gl(n|n)$ in
the fundamental representation. 
We choose the generators of $gl(n|n)$ as follows:
first $(2n)^2-2$ generators span a subspace of supertraceless and
traceless $2n \times 2n$ matrices and we denote them as $T_a$.
The remaining generators are the identity matrix
$I$, and matrix 
$J$, $J={\rm diag}(1, \dots ,1,-1, \dots, -1)$.
Also note, that $T_a$'s together
with the identity generate $sl(n|n)$, i.e. their commutators close
without $J$.

Generators $T_a$ get projected on generators of $psl(n|n)$ algebra
and we keep the same notations for them.  
The $T_a$'s do not generate $psl(n|n)$, as an identity appears
among commutators of $T_a$. However, if we ``ignore'' the identity part
of the commutators 
\eqn\comm{[T_a, T_b]=f_{ab}^{~~c} T_c + d_{ab} I~}
we would obtain the structure constants of $psl(n|n)$,
and the Jacobi identity is satisfied as one can easily check.
``Ignoring''
means subtracting identity to make the result traceless. Note that
the trace and supertrace operations are related to each other $Str(X)=Tr(XJ)$.

In our basis the $gl(n|n)$ metric looks like:
$$ \pmatrix{ g_{ab} &  0 \cr 
 \noalign{\smallskip} 0 & \matrix{ \scriptstyle 0 & \scriptstyle 2n \cr 
 \scriptstyle 2n & \scriptstyle 0 \cr }\cr }~. $$
The metric $g_{ab}$ is an invariant metric on 
$psl(n|n)$. 

\subsec{Comments on $GL(n|n)$ principal chiral model}

The group $GL(n|n)$ is not semi-simple and it has a {\it family} of 
invariant metrics $g^{(\mu)}_{ij}=Str(T_i T_j)+ \tilde \mu Str(T_i) Str(T_j)$.
This metric is non-degenerate for any value of $\mu$.
A simple calculation shows that at one loop a new term is generated
\eqn\actt{S={1 \over \lambda(\Lambda) ^2}\int {\rm Str} \big( |G^{-1} d
G|^2
\big)  +  \mu(\Lambda) \int | {\rm Str}( G^{-1} d G) |^2 ~~,}
which effectively introduces the cut-off dependence of the metrics 
$g^{(\mu)}_{ab}$.

To get the $SL(n|n)$ principal chiral model one just needs to restrict $G$
to lie in $SL(n|n)$ subgroup. There are two related issues
(i) the $SL(n|n)$ invariant metric is degenerate and
(ii) the $SL(n|n)$ principal chiral action has $U(1)$ gauge symmetry.
The appearance of this gauge invariance is quite remarkable and happens
only
for $SL(n|n)$. It acts as $G \rightarrow e^{\phi} G$. Under this symmetry
the current gets shifted $J \rightarrow J + d \phi$, but the action is 
still invariant because $e^{\phi}$ is {\it proportional} to the identity
matrix.   
Restricting oneself to a gauge invariant sector is the equivalent of 
making a quotient $PSL(n|n)=SL(n|n)/U(1)$. At the same time the problem (i)
also disappears, the $PSL(n|n)$ has an invariant non-degenerate metric.

The $gl(n|n)$ Lie algebra has a decomposition $ gl(n|n)=F_- \oplus B_0
\oplus F_+$, where $F_{\pm}$ are the subspaces of the lower and 
upper triangular odd matrices
and $B_0=gl(n) \times gl(n)$, the even part of the $gl(n|n)$.
Every element $Q_{ij} \in F_{\pm}$ is
nilpotent and $[F_-, F_+ ] \subset B_0$. The $GL(n|n)$
contains 
many operators that may play a role of the BRST operator. 
As we already explained, the left/right multiplication symmetry is broken 
by the choice of the vacuum (for simplicity we choose $G_0=I$). Still our
model is invariant with respect to conjugations.
As we will see, 
the $GL(n|n)$ is essentially a topological sigma model. 
For example, consider operator\ref\iss{J. M. Isidro and A. V. Ramallo, 
Phys. Lett. {\bf B340}, 48 (1994),
hep-th/9407152} 
\eqn\brs{Q=\sum_{i=1} ^{i=n} Q_{n+i ~j}~.}
The action of this operator corresponds to conjugation by the matrix
\eqn\matbr{\left( \matrix{
{\bf 1} & 0 \cr
\epsilon {\bf 1} & {\bf 1} }  \right)}
This operator is clearly nilpotent, but what is more important, the 
$GL(n|n)$ sigma model action is $Q$-exact! Indeed, if we represent the
current $J_{\mu}$
as $2 \times 2$ block matrix
\foot{It does not matter whether we choose the left or the
right current. Their transformation properties differ by sign.}
$$J_{\mu}=\left(\matrix{A_{\mu} & \chi_{\mu} \cr
\eta_{\mu} & B_{\mu} \cr} \right)$$
then the transformation properties are
\eqn\tran{\eqalign{
\delta_{\epsilon} A_{\mu}= \epsilon \chi_{\mu}~~,~~~\delta_{\epsilon}
B_{\mu} = \epsilon \chi_{\mu}\cr
\delta_{\epsilon} \chi_{\mu}=0 ~~,~~~\delta_{\epsilon} \eta_{\mu}= \epsilon
(A_{\mu}-B_{\mu})~.
}}
Now it is easy to see that the action \act\ is $Q$-exact
\eqn\exac{Tr(A_{\mu}A^{\mu})-Tr(B_{\mu}B^{\mu})+2 Tr(\chi_{\mu}
\eta^{\mu})=
[Q, Tr(\eta_{\mu} (A_{\mu}+ B_{\mu})]}
Similarly, the induced term in \actt\ is also $Q$-exact.
As the result, the computations of group invariant correlation functions
reduces to a classical problem. 

The BRST operator introduced in \brs\ turns out to be very useful. For
example, using the Q-cohomology technique one can prove the following
theorem:
if any Casimir operator has a non-zero eigenvalue on  
an irreducible representation $\Lambda$, 
then the super dimension of this representation is zero.  

The proof goes as follows: Consider the universal enveloping
algebra ${\cal U}({\cal G})$ (we would be interested in the 
case when 
${\cal G}=gl(n|n), sl(n|n)$ or $psl(n|n)$). The action of $Q$ on ${\cal G}$ 
can be lifted to an action on ${\cal U}({\cal G})$.  
Now, using the Poincare-Birkhof-Witt theorem one identifies
${\cal U}({\cal G})$ with 
$S({\cal G})=\oplus_{i=0} ^{i=\infty} S^i({\cal G})$, where 
$S^i ({\cal G})$ are symmetric powers of ${\cal G}$.
This identification is (i) an isomorphism of  ${\cal G}$-modules
and (ii) the space of invariants of  $S({\cal G})$ is mapped on the 
center in ${\cal U}({\cal G})$ (Casimirs). 
Now consider the 
action of $Q$ on $S({\cal G})$. It is clear that $Q:  S^i({\cal G})
\rightarrow S^i({\cal G})$ and one can define cohomology
$H^{*}(Q, S^i ({\cal G})$. Now, it can be shown that 
\eqn\cohh{
H^{*}(Q, S^i ({\cal G}))=
\left\{ \matrix{
0, & {\cal G}=gl(n|n)  & i>0 \cr
0, & {\cal G}=sl(n|n)  & i>1 \cr
0, & {\cal G}=psl(n|n) & i>2 } 
\right. }
For $sl(n|n)$ the cohomology
is non-zero only in degrees $n=0,1$ and is spanned by constants and $Q$.
For $psl(n|n)$ the cohomology is non-zero in degrees $n=0,1,2$
and the quadratic Casimir spans the cohomology in degree $n=2$.

As a  result all Casimir operators  are 
$Q$-commutators $\hat C^{(N)}=[Q, X^{(N)}]$ for $N>2$. 
The super dimension of any irreducible representation can be written as
\eqn\dim{{\rm sdim}({\Lambda})={\rm Str}(I)={1 \over C^{(N)}} 
{\rm Str}([Q, X^{(N)} ])=0~,}
where $C^{(N)}$ in the eigenvalue of the $N$th Casimir.
Now, for $psl(n|n)$ the quadratic Casimir is not $Q$-exact, but its
square is! Therefore, we can slightly modify our arguments in \dim\
by replacing ${\hat C}^{(2)}$ by its square.

This BRST-like symmetry can be used to compute some correlation functions
for $PSL(n|n)$ models using the localization technique. We will discuss
this issue in one of the next sections.

\newsec{Principal chiral field as a conformal theory}

\subsec{Background field method and conformal invariance.}

The simple way to show conformal invariance of the $PSL(n|n)$ principal
chiral model is based on the background field method
\ref\abbott{L. F. Abbott, Acta Phys. Polon. {\bf B13}, 33 (1982).}.
In this method the symmetries of the theory impose strong 
constraints on the form of the renormalized action, which is very useful.
We find that our theory is conformal to all orders in perturbation
theory.

First of all, the renormalization of the general non-linear sigma model was
studied extensively and the renormalization group flow for it was
interpreted as the flow in the space of metrics on the target manifold. 
In other words the shape and size of the manifold changes with the flow
\ref\friedan{D. H. Friedan, Ann. Phys. {\bf 163}, 318 (1958).}. 
The same topic was also studied using the background field method 
\ref\alvarez{L.~Alvarez-Gaume, D. Z.~Freedman and S.~Mukhi,
Ann. Phys. {\bf 134}, 85 (1981)}
\ref\mukhi{S.~Mukhi, Nucl. Phys. {\bf B264}, 640 (1986)}.
The main technical difficulty for the general case is the necessity to
expand the action in terms of ``linear'' fields, e.g. the Riemann normal
coordinates. In the case of the group manifold as a target space the
expansion simplifies dramatically and terms of any order can be written
explicitly. This will allow us to analyze all orders in perturbation
theory.

In the background field method we parameterize the quantum field $G(x)$ as
$g(x)G_0(x)$ where $G_0(x)$ is the classical background
and the field $g(x)$
describes quantum fluctuations. The ``linear'' field for the quantum
fluctuation is defined by $g(x)=e^{\lambda A(x)}$ where
$\lambda$ is inserted just to have a convenient normalization later. $A(x)$
is an element of the Lie algebra and transforms as a tangent vector at the
point $G_0(x)$. In terms of these fields the (right) current of the model
becomes
\eqn\cur{J^G_\mu \equiv G^{-1}\del_\mu G =  G_0^{-1}\del_\mu G_0 + 
 G_0^{-1}j^A _{\mu}G_0= 
  J^0 _\mu+ G_0 ^{-1} j^A _{\mu} G_0~,}
where we denoted the background current by $J^0$ and the current
corresponding to quantum fluctuations as 
$j_{\mu} ^A=e^{-\lambda A}\del_\mu e^{\lambda A}$. 
The action is then
\eqn\Laggr{S[G]=S[e^{\lambda A}]+S[G_0]+{ 1 \over 2 \pi \lambda^2} \int Str
\bigl( j^A _{\mu} (\del^\mu G_0) G_0^{-1} \bigr) 
}
The current $j^A$ written in terms  
of $A$ has the following expansion:
\eqn\currentA{\eqalign{j^A_\mu \equiv e^{-\lambda A}\del_\mu e^{\lambda A}
&= 
           \sum_{n=1}^{\infty}{ \lambda^n \over n!} 
                  [.. [[\del_\mu A, A ],A]\ldots A ] = \cr
               &= \lambda \del_\mu A + { \lambda^2 \over 2 } [\del_\mu A, A
               ] + 
          { \lambda^3 \over 3! }[[\del_\mu A, A ],A] + \ldots 
}}
We will need the polynomial expansion of \Laggr\ in terms of $A$:
\eqn\LagrnA{\eqalign{  L(e^{\lambda A})& ={1 \over 2 \pi}
  \sum_{n=1}^{\infty} {\lambda^{(2n-2)} \over (2n)!} Str\bigl( 
       [.. [[\del_\mu A, A ],A]\ldots A ] \del^\mu A \bigr) = \cr
  &=  {1 \over 4 \pi} \big(
Str( \del_\mu A \del^\mu A ) + 
  { \lambda^2 \over 12 } Str( [[\del_\mu A,A],A] \del^\mu A ) + 
  \ldots \big)
}}

Putting it all together we find that our Lagrangian 
for the quantum field $A(x)$ in the background
$G_0(x)$ contains the following terms: an $A$-independent part which
is the Lagrangian for the background field itself, free-field kinetic
energy $Str( \del_\mu A \del^\mu A )$ and the interaction terms of two
kinds. First, the interaction terms which involve the background current
and single derivative of $A$ and second, the terms that appear in the
expansion
\LagrnA, which contain two derivatives of $A$. 
An important point to notice is
that all interaction vertices are built from structure constants of
$psl(n|n)$. Schematically, each interaction vertex can be represented 
as a tree diagram, shown in Fig. 1. These diagrams describe the 
``color structure'' of the interaction vertices.
Each three-vertex in the picture represents a $psl(n|n)$
structure constant $f_{abc}$ and the dashed lines correspond to the
contraction of indices with the $psl(n|n)$ invariant metric $g_{ab}$ but no
propagator insertion.

\bigskip


\centerline{\epsfxsize 4.truein \epsfbox{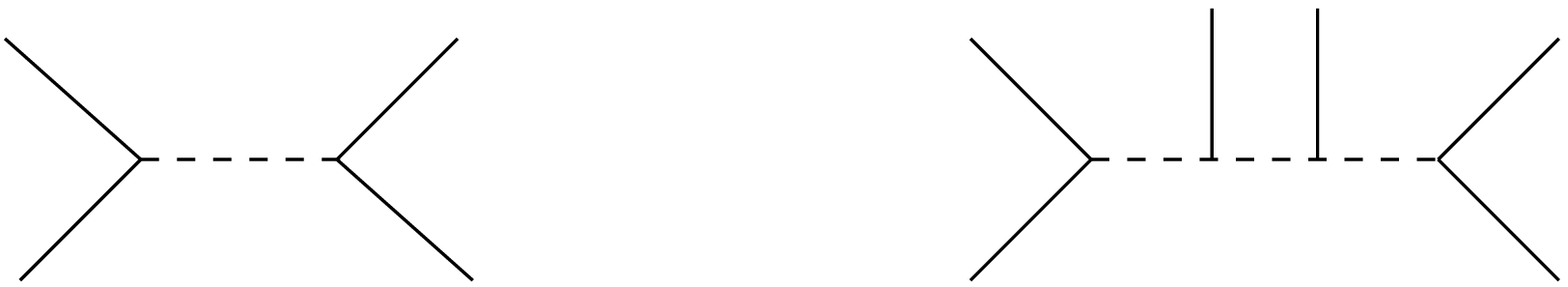}}

\noindent{\ninepoint\sl \baselineskip=8pt {\bf Fig.1}
{\rm Typical
interaction vertices}}

\bigskip

To compute effective action for the background field $G_0 (x)$, one has to
evaluate all 1PI diagrams with external background lines only.  The typical
diagram is represented in Fig. 2.  To find the beta function we have to
renormalize UV divergent diagrams.  There are of course IR divergent
diagrams, but we assume that they have been regularized as discussed
before, say, by including a small mass term.  Notice that the structure of
interaction vertices is such that the background current appears in
vertices with a {\it single} derivative of $A$. Therefore, by power
counting, all primitively divergent diagrams contain no more than two
external lines of the background field.  Those diagrams with no background
external lines at all will lead to renormalization of the action for the
quantum field $A$.  The resulting wave function/vertex renormalization for
it would not affect our conclusions because there are no external
$A$-lines. By choosing the renormalization scheme that preserves the symmetry
of the problem, one also ensures that the group structure of renormalized
vertices does not change.

\bigskip


\centerline{\epsfxsize 2.8truein \epsfbox{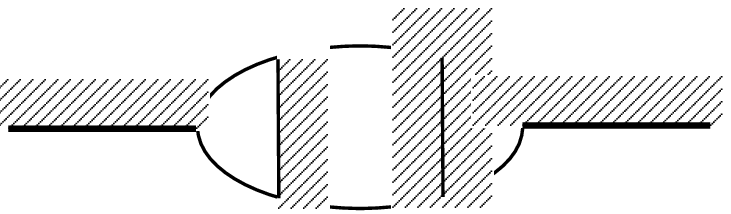}}

\noindent{\ninepoint\sl \baselineskip=8pt {\bf Fig.2}: {\rm Example of a
diagram with two external background lines}}

\bigskip

Now we will demonstrate that the diagrams with one or two background
external lines are identically equal to zero, even before one computes
the momentum integral. All these diagrams vanish because of the 
group factors. 
Consider first the diagrams with a single external background line. 
As we mentioned, we represented the interaction vertices as tree
diagrams constructed from three-vertices (structure constants). Let us take
the 3-vertex where the external line enters and pull it out of the 
diagram (see Fig. 3a). The rest of the diagram can be represented as a blob
with two external lines. Its group structure is given by a second rank 
{\it invariant} tensor. There is only one such tensor -- the metric 
and its contraction with the structure constants vanishes.

\bigskip


\centerline{\epsfxsize 5.1truein \epsfbox{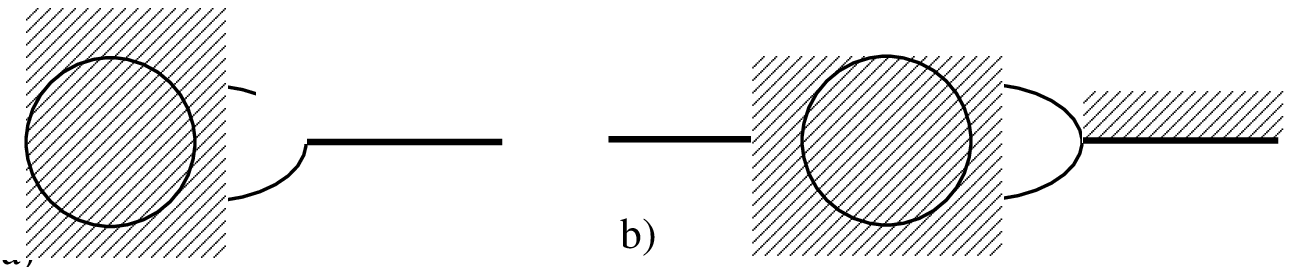}}

\noindent{\ninepoint\sl \baselineskip=8pt {\bf Fig. 3} }

\bigskip

Next let us consider the diagrams with two external background lines and 
do the same trick. Let us pull out the 3-vertex where the external 
line enters. The rest of the diagram is now a blob with 
three external lines, one of them is the background field line and 
the other two  are contracted with the structure constants (see Fig. 3b).
This blob is also an invariant tensor. 
Luckily enough, $psl(n|n)$ has only one invariant rank 3 tensor -- 
structure constants. This can be proved as follows: all rank 3 invariant 
tensors come from the $gl(n|n)$. For $gl(n|n)$ there are $6$ invariant 
rank 3 tensors but only one survives the reduction to $psl(n|n)$.
As a result, the whole contraction is proportional to the metric times
the dual Coxeter number $f_{abc}{f_d}^{bc} = C_V g_{ad}$. However, that
number vanishes for the $psl(n|n)$ groups
\foot{There is another group for which dual Coxeter number vanishes --
$osp(2n+2|2n)$. It is possible that the $osp(2n+2|2)$ sigma model 
is also conformal}! 

What we just showed is that there are no divergent diagrams 
for the effective action and 
therefore the coupling $\lambda$ is not renormalized to all orders in
perturbation theory. Hence the $psl(n|n)$ principal chiral model is 
perturbatively conformal.

\subsec{Perturbation theory.}

In this section we present an analysis of the perturbation theory for
the $PSL(n|n)$ principal chiral model. We find that some correlation
functions can be computed exactly to all orders in perturbation theory, in
an
analogy to the calculations in the background field method. We will also
prove conformal invariance in a different way. 

The very definition of the model requires breaking of the $G_L \times G_R$
invariance in order to get rid of the fermionic zero modes. We choose a
``false'' vacuum $G(x)=G_0$ and, if working in an infinite volume, 
add a small potential term making  $G_0$ a true ground
state. This is exactly what is necessary to build a perturbation theory.
The action in the ``background'' of the $G_0$ vacuum is given by 
\Laggr\
plus the potential term that we choose as $m^2 Str(A^2)$
\eqn\Spert{\eqalign{S_{pert}[A;G_0]&=S[e^{\lambda A}] - 
{1 \over 4 \pi}  \int d^2 \sigma \, m^2 Str( A^2 )  = \cr
  &={1 \over 4 \pi} \int d^2 \sigma \, Str( \del_\mu A \del^\mu A - m^2 A^2
  ) + S_{int}
}}
Although this action explicitly breaks $G_L \times G_R$ symmetry it is
still invariant under the subgroup which leaves $G_0$ invariant. This
remaining symmetry acts on $A$ simply by conjugation $A \to h^{-1} A h$ and
on $G(x)$ by $G(x) \to h G(x) G_0^{-1} h^{-1} G_0$. We will explore
consequences of this symmetry later on.

Once again, the most important thing to notice about \Spert\ is that all
the interaction vertices in $S_{int}$ are built from the structure
constants of $psl(n|n)$ in exactly the same fashion as we found in the
background field method (See Fig. 1). In particular, all vertices have the
group structure of tensors invariant under the remaining symmetry, i.e.,
tensors invariant under the conjugation by the elements of the Lie algebra
of $psl(n|n)$.

Now suppose we want to compute a correlation function which is manifestly
invariant under the left {\it and} right multiplication symmetry $G_L
\times G_R$. As explained, we expect to trust the answer computed in the
broken symmetry ``phase'' and we also know that in perturbation theory the
result is IR-finite. Invariance under $G_L \times G_R$ symmetry
implies that the correlation functions are invariant
under the symmetry remaining in the perturbation theory -- 
the conjugation of
$A$. It means that in every order in $A$ the expression inside the
correlation function can be written as a product of $A$'s and their
derivatives with all their group indices contracted with an invariant
tensor:
$$ \langle (\del)A^{a_1}(x_1) (\del)A^{a_2}(x_2) \ldots (\del)A^{a_n}(x_n) 
\, d_{a_1 a_2 \ldots a_n}  \rangle 
$$
Here $d_{a_1 a_2 \ldots a_n}$ is an invariant $n$-tensor. Then the group
structure of any Feynman diagram contributing to such a correlation
function, even before the momentum integration, can be represented as
follows.  It is a collection of invariant tensors of $psl(n|n)$, one coming
from the above expression and others from $S_{int}$, and all their indices
are contracted with the help of the inverse metric on $psl(n|n)$. Most
importantly, any subdiagram is then also an invariant tensor.  The rank of
this tensor is given by the number of legs connecting it to the rest of the
diagram.

This allows us to do the same trick as before. Consider any interaction
vertex inside a given Feynman graph. Take any three-vertex in it and pull
it out. (see Fig. 4) The rest of the diagram is an invariant tensor with
three indices.  There is only one such tensor for $psl(n|n)$ - structure
constants.  The pairwise contraction of all indices of two three-tensors is
proportional to the dual Coxeter number $f^{abc} f_{abc}=C_V \delta^a
_a=0$. Thus all the diagrams without external lines and with at least one
interaction vertex vanish. The remaining diagrams represent the calculation
made in the free theory, $S_{int} =0$. Moreover, the $A$'s are effectively
commuting variables in such calculations since commutators introduce more
of the structure constants in the graph and their contributions vanish.

\bigskip


\centerline{\epsfxsize 1.1truein \epsfbox{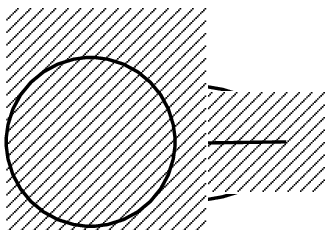}}

\noindent{\ninepoint\sl \baselineskip=8pt {\bf Fig. 4} }

\bigskip

It may also happen that the invariant tensor $d_{a_1 a_2 \ldots a_n}$
entering the correlation function would itself contain $f_{abc}$. Such
correlation functions all vanish, by the same reasoning.

Let us illustrate the preceding discussion with some examples. Consider
$G_L$ and $G_R$ invariant bilinears built from the (right) currents:
$Str(J_z J_z)$ and $Str(J_z J_{\bar z})$. We can compute expectation values
of strings of such operators. The currents themselves are complicated
expressions in terms of $A$, given in \currentA. Fortunately, all the terms
but one contain commutators. Therefore the invariant tensors appearing in
the correlation functions with those terms all contain structure constants.
Thus the only contributing terms are coming from $\del A$ and $\bar
\del A$. The calculation reduces to computing expectation values of strings
of $Str(\del A \del A)$ and $Str(\del A {\bar \del}A)$ in the free theory
with commuting $A$'s. For example:
\eqn\opp{\langle Str(J_z(z,{\bar z}) \, J_z(z,{\bar z})) \, 
         Str(J_z(0,0) \, J_z(0,0)) \rangle = -{4 \lambda^4 \over z^4}
}
Analogously we find
$$
\eqalign{\langle Str(J_z(z,{\bar z}) \, J_{\bar z} (z,{\bar z})) \, 
         Str(J_z(0,0) \, J_z(0,0)) \rangle =  0}
$$
$$
\eqalign{\langle Str(J_z(z,{\bar z}) \, J_{\bar z} (z,{\bar z})) \, 
         Str(J_z(0,0) \, J_{\bar z}(0,0)) \rangle = 
                   - { 2 \lambda^4 \over |z|^4 }
}
$$

Given these correlation functions we can present another proof that our
theory is conformal at the quantum level.  The classical theory is
conformal and therefore the $z - \bar z$ component of the stress-energy
tensor is equal to zero. Quantum mechanically, there could an
anomaly. However the symmetries of the problem restrict its possible form
\ref\wit{E. Witten and 
Y. Goldschmidt, Phys. Lett. 91B, 392 (1980)}.
It can only be that $T_{z \bar z}= \alpha Str (J_z J_{\bar z})$, where
$\alpha$ is a constant, possibly dependent on the scale.  Now we want to
show that $\alpha=0$.  Following Zamolodchikov
\ref\zam{A. B. Zamolodchikov, JETP Lett. {\bf 43} 730 (1986)}
we introduce functions $G(z \bar z)$ and $H(z \bar z)$
(see also \ref\polch{J. Polchinski, Nucl. Phys. {\bf B303}, 226 (1988)})
\eqn\zamm{\eqalign{G(z \bar z)=4z^3 \bar z \langle 
T_{zz}(z, \bar z) T_{z \bar z}(0)  \rangle \cr
H(z \bar z)=16 z^2 \bar z^2 \langle 
T_{z \bar z}(z, \bar z) T_{z \bar z}(0)  \rangle
}}
The conservation of stress-energy tensor and rotational invariance imply
that
$$4 \dot G - 4 G + \dot H -2 H=0~,$$ 
where dot denotes the derivative with respect to $r^2=z \bar z$.  Now,
taking into account that $T_{zz}=-{1 \over 2 \lambda^2}Str(J_z J_z)$ we
conclude that $\dot H-2H=-32 \alpha^2 \lambda^4=0$. This means that $T_{z \bar z}$ is
actually zero quantum mechanically in perturbation theory. Notice, that we
never used any information about whether the theory is unitary or not.

As a result we find that our theory is conformal. It also follows from 
\opp\ that the central charge is equal to $(-2)$ for both left and 
right Virasoro algebras, indeed
\eqn\sttr{\langle T(z) T(w) \rangle ={-1 \over (z-w)^4}.}

Finally, we can extend our results on correlation functions that can be
computed exactly. Suppose that we want to compute a correlation function
which is invariant only under {\it one} of the $G_L$ or $G_R$ and is an
invariant tensor under the action of the other. It turns out that if this
tensor is of the rank two, we can still do computations as in the free
theory. There are no invariant tensors of rank one. The only rank-two
tensor is the metric, so the correlation function is proportional to it. To
find the coefficient of proportionality we can contract those two
indices with the inverse metric, which gives $c$ times $g_{ab} g^{ab} =
-2$. The correlation function thus becomes invariant and can be computed in
the free theory just as before.

\subsec{Some examples of perturbative calculations}

The $psl(n|n)$ is a conformal theory and therefore all fields can be 
decomposed into 
represenations of the Virasoro algebra. The current $J_{\mu}$ is an
interesting 
example. It satisfies the equation of motion $\partial_{\mu} J^{\mu}=0$ and 
is not holomorphic. On the other hand one can suggest that the current 
component
$J_z$ (or $J_{\bar z}$) is a $(1,0)$ (or $(0,1)$) conformal 
primary field. 
Notice that the theory in question 
is not {\it unitary} and therefore the zero norm states
(such as $\bar \partial J_z$) do not decouple.
The currents are invariant with
respect to either left or right multiplication symmetry and therefore
one can analyze multipoint correlation functions of left (or right)
currents using perturbation theory.
For example, the two point
correlation function of currents 
is known exactly
\eqn\tt{ \langle J_a(x) J_b(y) \rangle= {\lambda^2 g_{ab} \over (x-y)^2}}
As was explained before, all loop 
diagrams (in other words, diagrams with interaction vertices)
that may contribute to this calculation have two external 
lines and therefore vanish  identically. 
The 3-point correlation function is almost uniquely fixed by conformal
invariance 
\eqn\thr{\langle J_a(x)J_b(y) J_c(z) \rangle
={3 \over 2}{\lambda^4 f_{abc} \over (x-y)(x-z)(y-z) }~,}
modulo the scale coefficient that we claim is exactly equal to $3 \lambda^4
/2$.
Let us present the arguments that this is an exact answer.
Consider the Feynman diagram that contributes to \thr.
It contains three-vertices coming both from the expansion of the currents
\currentA\ and from the interaction vertices. Diagrams containing only one
three-vertex come from the second term in the current and give \thr. We
claim that any Feynman diagram containing two or more three-vertices
vanishes. To show this let us pull out the three-vertex where the external
line enters. The rest of the diagram is a blob with four external lines
(see Fig. 5). The group structure of this blob is a rank 4 invariant
tensor. Let us contract any of its two indices. The resulting graph
contains a three-vertex inside and vanishes by our previous arguments.
Therefore the blob is a traceless rank 4 tensor. There are only five
independent rank 4 invariant tensors: $g_{ab}g_{cd}$ and two other tensors
with indices being permuted (three in total) and ${f_{ab}}^r f_{cdr}$ and
${f_{ac}}^r f_{bdr}$. Only three combinations of those are traceless:
${f_{ab}}^r f_{cdr}$, ${f_{ac}}^r f_{bdr}$ and $g_{ab}g_{cd}+g_{ac}g_{bd}+
g_{ad}g_{bc}$. Luckily, for all of the traceless tensors the contraction of
any two indices with the structure constant $f_{abc}$ produces zero. This
explains why \thr\ is exact.

\bigskip


\centerline{\epsfxsize 2.6truein \epsfbox{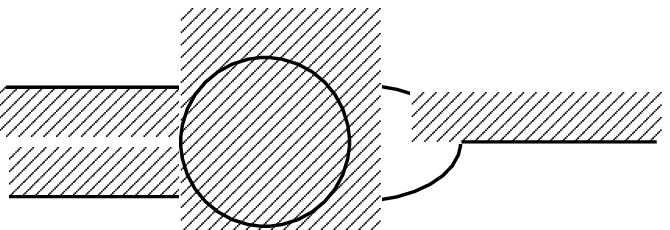}}

\noindent{\ninepoint\sl \baselineskip=8pt {\bf Fig. 5} }

\bigskip

The higher point correlation functions are nontrivial and have  
complicated $\lambda$ dependence. For example
on the general grounds one can conclude that  
$$\langle
J_{a_1}(z_1)J_{a_2}(z_2)J_{a_3}(z_3)J_{a_4}(z_4) \rangle
=\lambda^4
       \Pi_{i<j} (z_i-z_j)^{-2/3}\sum_k t^{(k)}_{a_1...a_4}f_k(x) g_k(\bar
       x),$$
where $t^{(k)}$ are possible tensor structures and
$x$ is an anharmonic ratio. This correlation 
function is clearly not holomorphic. Notice, that $\lambda^4$
is just a normalization factor. The functions $f_k(x), g_k(\bar x)$
also have a non-trivial dependence on $\lambda$.

\subsec{Comments on the WZW term}

In the previous sections we proved a remarkable statement --
the principal chiral field for $psl(n|n)$ is a conformal theory. 
There is another well known way to associate a conformal field theory with
any group (allowing a non-degenerate invariant second rank tensor) -- 
WZW theory 
\ref\conwz{E. Witten, Comm. Math. Phys. {\bf 92} 445 (1984)} 
\ref\kz{V. G. Knizhnik and A. B. Zamolodchikov, 
Nucl. Phys. {\bf B247}, 83 (1984)}. 
The action for WZW theory is just a combination of principal
chiral action and a WZW term.
The theory is conformal (and possesses the affine symmetry algebra)
when the coefficient in front of the  principal
chiral action ${1 \over \lambda^2}$ coincides with the coefficient $k$
in front of the WZW term.
If these coefficients are different the theory flows, or in other words it
is not conformal. 

For $psl(n|n)$ group the story is different. The theory is conformal 
for any values of $k$ and $\lambda$ 
\eqn\awzw{S_{wzw}[G]={1 \over 4 \pi \lambda ^2}\int {\rm Str} \big( |G^{-1}
d
G|^2  \big)     + k \Gamma_{wzw} [G]~.} 
The proof of this statement is almost identical to the one we just gave for 
$psl(n|n)$ principal chiral field. It is enough to say that all vertices
introduced by WZW term have the group structure which can be described by the
similar tree-like diagrams (see Fig. 1). 

As a result we have a two parameter family of conformal field theories,
parameterized by two charges $k$ and $\lambda$. For compact groups
(or for groups having non-trivial $H^3$) the 
coefficient $k$ is 
quantized, then it is natural to think about parameter $\lambda$ as an
exactly marginal perturbation. At point $k=1/\lambda^2$ the global
left/right multiplication symmetry is enhanced and the theory is invariant
under the left/right current algebra. As we see, the theory still
possesses a huge chiral algebra for any value of $\lambda$, 
which is an extension of Virasoro algebra.

Let us see  what is the physical meaning of the coefficient $k$ in front of the
WZW term and the radius of sigma model 
$R={1 \over \lambda}$ in the $AdS_3 \times S^3$ applications.
The simple way to realize $AdS_3 \times S^3$ geometry is to consider
configuration
of $Q$  NS 5-branes and $p$ fundamental strings 
\ref\mar{A. Tseytlin and M. Cvetic, Phys. Rev. {\bf D53} 5619 (1996), 
hep-th/9512031 \hfil\break 
A. Tseytlin, Mod. Phys. Lett. {\bf A11} 689 (1996),
hep-th/9601177}. 
This configuration can
be described by WZW model 
(see for example \kut) 
with ${1 \over \lambda^2}=k_{WZW}=Q$. Now, performing the S-duality
transformation, one can map this configuration on $(Qa,Qc)$ 5-branes and 
$(pa,pc)$ strings (the second integer indicates the number of D-objects). 
Then the radius of $AdS$, measured in Einstein frame, remains invariant
under the S-duality transformation. As a result we end up 
with  $1/\lambda^2=R^2= \sqrt{(Qa)^2+(Qc)^2 g^2}$,
where $g$ is the string coupling constant and $Qc$ counts the number of $D$ 
5-branes.
The coefficient in front of the WZW term counts the NS 5-brane 
charge and is equal to $Qa$.
It was noticed in \bvw\ that for $1/\lambda^2 < Qa$ the theory becomes
ill defined.   

\newsec{Comments on Localization}
  
In this section we give some non-perturbative arguments about $PSL(n|n)$
principal chiral model. It will add strength to the perturbative
calculations of the previous section. 

Recall that the symmetry groups of our model, $G_L$ and $G_R$, are actually
supergroups and as such posses a number of fermionic generators. Those
generators act without fixed points and because of that we have to break the
$G_L \times G_R$ symmetry and define the model as a limit of the theory
with a unique vacuum, $G_0$. The symmetry that remains in that latter
theory is the diagonal subgroup of $G_L \times G_R$. 
It acts by conjugation (we chose $G_0 = \bf 1 $ for simplicity)
\eqn\adjaction{
G(x) \to U G(x) U^{-1} , \quad \quad U \in PSL(n|n)}
Now, $PSL(n|n)$ has $2n^2$ fermionic generators which can be
considered as BRST charges. We split them into two subgroups mentioned
earlier, $F_+$ and $F_-$
$$ F_+ = \pmatrix{ {\bf 1} & \chi_+ \cr 0 & {\bf 1}} \quad \quad
   F_- = \pmatrix{ {\bf 1} & 0 \cr  \chi_- & {\bf 1}}
$$ 
where $\chi_{\pm}$ are fermionic $n \times n$ matrices. Both $F_+$ and
$F_-$ are abelian subgroups of $PSL(n|n)$.

The $G_L \times G_R$ invariant correlation functions are at the same time
BRST invariant and one can localize the path integral to the arbitrarily
small neighborhood of a set fixed by the BRST action
\ref\local{E.~Witten, Nucl. Phys. {\bf B371}, 191 (1992)}.
In our case the generators of the $F_\pm$ can all be considered as BRST
charges and the only point fixed by them is the vacuum state $G(x)= \bf 1$.
Thus to evaluate the path integral in the broken symmetry ``phase'' with
$G_L \times G_R$ invariant observables it is sufficient to restrict
ourselves to the infinitesimal neighborhood of the vacuum, i.e. the
Gaussian approximation is exact. This confirms what we found earlier by
analyzing perturbation theory. Namely, in computing $G_L$ and $G_R$
invariant correlation functions we can set $S_{int}$ to zero.  In the case
of correlation functions invariant only under a subgroup of $G_L \times
G_R$ it is possible to localize the path integral on a bigger set.

The above arguments can also be made in the case of $GL(n|m)$ principal
chiral models. However, it doesn't make those theories conformal. An
important ingredient for doing exact computations in perturbation theory is
missing.  In the $PSL(n|n)$ case any invariant diagram containing the
three-vertex vanishes, even if the vertex comes not from the action but
from the expansion of the operators.  In the case of $GL(n|m)$ this is no
longer true and although the calculations can be made in the free theory,
the infinite number of terms arising, for instance, from the expansion of
$J$ would all contribute. 

\newsec{Chiral algebra}

As we just demonstrated, the $PSL(n|n)$ principal chiral model is a
conformal
field theory. The central charge is equal to $c=-2$ and 
is independent on $\lambda$.
As we will see, the parameter $\lambda$ is very similar to the radius
parameter 
in the free theory on a circle.

It turns out that the chiral algebra of the 
theory is not just Virasoro algebra, 
but much bigger.   
Classically, any chiral model contains an infinite set of holomorphic
currents. To be more precise, for every invariant symmetric
tensor $t_{a_1...a_n}$ one can construct a holomorphic current
\eqn\holcur{W_{[t]}={1 \over n!} t_{a_1 a_2 ...a_n} J^{a_1} J^{a_2}...J^{a_n}}
It is irrelevant whether we use the left or the right current in the 
definition of $W_{[t]}$.
The equation of motion 
\eqn\mot{\partial_{\mu} J^{\mu}=\partial_{\bar z}J_{z}+ \partial_{z}
J_{\bar
z}=0~,}
being combined with the flatness condition
\eqn\flat{\partial_{\bar z}J_{z}- \partial_{z} J_{\bar z} +[J_{\bar z},
J_{z}]=0}
yields the relation
\eqn\imp{\partial_{\bar z}J_{z}+{1 \over 2}[J_{\bar z}, J_{z}]=0 }
It is easy to see that the commutator in the last equation does not 
contribute to $\partial_{\bar z} W_{[t]}$ and 
as a result, $W_{[t]}$
turns out to be holomorphic at the classical level (see for 
example \wit). 
For an arbitrary group $G$ the holomorphicity of the currents $W_{[t]}$ is
destroyed
by quantum corrections. 
However, the case of $PSL(n|n)$ group is very remarkable and we believe
that the currents
$W_n$ remain holomorphic.

The construction of totally symmetric invariant tensors is given in the 
Appendix. Here we just show one example.
Consider $SL(n|n)$ invariant tensor corresponding to normalized trace 
$d_n(X_1, X_2,...X_n)=(1/n!) Str(X_1X_2...X_n)$. 
Under $U(1)$ transformation 
$G \rightarrow e^{\phi} G$ it
transforms as follows: $\delta d_n(...)=
(1/n) \partial \phi \sum_i d_{n-1}(...)$. 
Given this, it is easy to see that for every $n$ one can construct an
invariant tensor of degree $2n$
\eqn\invten{t_{2n} \sim {1 \over 2} d_n^2 + \sum (-1)^i d_{n-i} d_i}
Each $psl(n|n)$ invariant tensor $t_n$ gives rise to a holomorphic 
current $W_{[t]}$ that remains holomorphic even at the quantum level.
One can check this by analyzing the perturbation theory
(we checked the first few non-trivial orders in perturbation theory).
The field $W_{[t]}$ is a conformal primary field of dimension
$\Delta={\rm rank}(t)$ if tensor $t$ is traceless. 
This can be always achieved by subtracting the traces.

The fields $\{ W_{[t]} \}$ generate the chiral algebra of the theory.
The first field is the stress energy tensor $T=W_2$, the next ones are
$W_{[t_6]}$
and $W_{[t_8]}-T^4/480$.
For $psl(2|2)$, all other fields $W_{[t]}$ are in a sense ``dependent''. 
This dependence is non-polynomial. 
For example 
$$W_{[t_8]} T^2 -{2 \over 5} W_{[t_6]} ^2 - {1 \over 720} T^6+...=0~,$$
 where the dots represent possible quantum corrections. 
This chiral algebra is an example of a  $W$-algebra. 
The OPE of $W_{[t_6]}$ with 
itself has the following structure
\eqn\oppe{W_{[t_6]}(z) W_{[t_6]} (w)=
{C_{66} ^8 \over (z-w)^4} [W_{[t_8]}]+ 
{ C_{66} \over (z-w)^8 } [U]~.}
The brackets $[X]$ denote the conformal block of operator $X$.
The field $U$ is given by a combination of square of stress 
energy tensor and its second derivative.
In general, for $psl(n|n)$ theory there are $(n-2)$ independent operators.  
The algebra generated by these currents is quite complicated and it would
be 
nice to have some explicit description of this algebra.
For example, it is plausible that this chiral algebra can be obtained by 
Hamiltonian reduction from $psl(n|n)$ affine algebra. 
This representation, being
constructed, should be very helpful for representation theory.

\newsec{Physical operators and Kac representations}

\subsec{Sigma model operators}

Let us describe operators that appear in the quantization of the $PSL(n|n)$
sigma model. All fields can be classified by representations of the
symmetry group $G_L \times G_R$, or in other words, by a pair of weights
$(\Lambda_L, \Lambda_R)$
\foot{Although not all the representations are of the highest
weight type, we would not have the need for those in our discussion.}. 
Let us denote by $V_{(\Lambda_L, \Lambda_R)}$ a
primary field of the chiral algebra. Observe that all the $W_{[t]}$ can be
constructed either from the left current $J_L$, or from the right current
$J_R$.  Therefore, in order to avoid the contradiction, we have to assume
that the weights $(\Lambda_L, \Lambda_R)$ are not arbitrary but such that
all the Casimir operators have the same eigenvalues for both $\Lambda_L$
and $\Lambda_R$ representations. 

In this paper we discuss only highest/lowest weight representations.
For generic weight $\Lambda$ the
Kac representation (see definition in the Appendix) is irreducible, 
but for certain weights $\Lambda$ it turns out to be reducible (atypical).
The resulting {\it irreducible} representation turns out to be smaller
and we call it a short representation. These representations will play in
important role in the $AdS/CFT$ duality. They correspond to chiral
multiplets in the boundary theory.   

One can easily construct operators corresponding to the weights $(\Lambda,
\Lambda'=\Lambda)$ (both left and right representations are the same) of
$PSL(n|n)$.  For our group every Lie algebra representation gives rise to
the group representation. 
Let $T^a$ be the generators of the Lie algebra in the representation
$\Lambda$ and let $A^a$ be the corresponding coordinates on the Lie
algebra.  Then we propose $V_{(\Lambda, \Lambda)}=e^{\lambda A^a T^a}$ as a
candidate for the {\it conformal primary} field.  Classically, this field
has zero dimension, but it acquires an anomalous dimension. In the leading
order it is equal to $\Delta= \bar \Delta=\lambda^2 C_{\Lambda}/2$, where
$C_{\Lambda}$ is the eigenvalue of the quadratic Casimir. We conjecture
that this expression for the dimension is exact to all orders in $\lambda$.

Let us discuss the correlation functions of these operators.  We first
consider two-point correlation function of a group invariant combination.
In order to construct a group invariant combination of two
representations $\Lambda_1,~ \Lambda_2$, one needs to have a way of pairing
two representations (``contracting indices'')
\foot{In other words, one has to choose elements in
$Hom(\Lambda_1 ,\Lambda_2^*)$ and $Hom(\Lambda_1^* , \Lambda_2)$.}.
For two irreducible representations this pairing exists only if
$\Lambda_1=\Lambda_2 ^*$ and in this case $(V_{(\Lambda^*,
\Lambda^*)})^k _j=(V_{(\Lambda, \Lambda)} ^{-1})^k _j$. Then 
the pairing is
just the supertrace of the product of two matrices.  As we know, an
invariant two point correlation function can be easily computed to all
orders in perturbation theory
\eqn\allor{ \langle Str (V_{(\Lambda, \Lambda)}(z) 
V_{(\Lambda^*, \Lambda^*)}(w)) \rangle={{\rm sdim}({\Lambda}) \over |z-w|^{2
\lambda^2 C_{\Lambda}}}}
Unfortunately, this result is not too exciting. The superdimension is equal
to zero unless $C_{\Lambda}=0$. Therefore, this correlation function
differs from zero only for short representations.  One may try to consider
correlation functions of non-invariant operators, for example
$V_{(\Lambda, \Lambda)}(z) \cdot V_{(\Lambda^*, \Lambda^*)}(w)$ (no
supertrace). This correlation function is invariant under the right
multiplication and so can be computed around any $G_0$ vacuum. On 
general grounds one expects the answer to be of the form
\eqn\nonin{ \langle (V_{(\Lambda, \Lambda)})^i _k (z) 
(V_{(\Lambda^*, \Lambda^*)})^k _j(w) \rangle=
{ N \delta_j ^i \over |z-w|^{2 \lambda^2 C_{\Lambda} } }~.}
To compute the normalization factor $N$ one may project on invariant
subsector by taking a trace. As a result one gets 
$N {\rm sdim}({\Lambda})={\rm sdim}({\Lambda})$. 
Therefore it implies that $N=1$, if
${\rm sdim}({\Lambda}) \neq 0$. In the case when ${\rm sdim}({\Lambda}) =0$ we
can not justify that $N=1$ using this kind of argument. Still,
perturbation theory predicts that $N=1$.

The correlation function of matrix elements $(\langle V_{(\Lambda,
\Lambda)})^i _k (V_{(\Lambda^*, \Lambda^*)})^k _j \rangle$ can not be
computed in perturbation theory.  However, from the transformation
properties of the correlation function we expect that
\eqn\nonbad{ \langle (V_{(\Lambda, \Lambda)})^i _k (z) 
(V_{(\Lambda^*, \Lambda^*)})^l _j(w) \rangle= \delta^i _j \delta^l _k
(-1)^{\bar k^2} f(z,w) }
where $\bar k$ is the parity of the vector $k$ in the representation
$\Lambda$.
To compute $f(z,w)$, one can just evaluate the trace over indices $k,l$
producing a simple  relation 
${\rm sdim}({\Lambda}) f(z,w)= 1/|z-w|^{\lambda^2 C_{\Lambda}}$. 
For short representations ${\rm sdim}({\Lambda}) \neq 0$ 
and at the same time $C_{\Lambda}=0$ (see section 2.2) and we obtain 
$f(z,w)=1/{\rm sdim}({\Lambda})$. 
For generic representations these arguments fail and we
do not know how to compute correlation functions of matrix elements.

In fact, we can also discuss the multipoint correlation functions of group
invariant combination of operators. These correlation functions can be
computed to all orders in perturbation theory. Again, to define the 
three point correlation functions, one needs Clebsch-Gordan
coefficients in order to contract indices.  Saying it 
differently, one has to
fix two invariant tensors $\phi$ and $\chi$
\foot{In other words,
$\phi \in Hom(\Lambda_1 \times \Lambda_2 , \Lambda_3 ^*)$ and $\chi \in
Hom(\Lambda_1^* \times \Lambda_2^*,
\Lambda_3)$ are non-zero if representation $\Lambda_3$ appears in
decomposition $\Lambda_1 \times \Lambda_2$.}.
When the choice of $\phi$ and $\chi$ is unique, they are necessarily proportional
to Clebsch-Gordan coefficients (as in the case of the classical groups).
This choice fixes the gauge invariant combination.  Now,
going through the similar computations one finds that
\eqn\thre{\eqalign{ \langle   \phi \otimes \chi 
\big( V_{(\Lambda_1, \Lambda_1)}(z)  
      V_{(\Lambda_2, \Lambda_2)}(w) 
      V_{(\Lambda_3, \Lambda_3)}(x) \big) \rangle= \cr 
={ \langle \phi ,\eta \rangle
\over |z-w|^{\gamma_{12}}|z-x|^{\gamma_{13}}
|w-x|^{\gamma_{23}} }~,}}
where $\gamma_{12}=2 \lambda^2 (C_1+C_2-C_3),\gamma_{13}=2 \lambda^2
(C_1+C_3-C_2),
\gamma_{23}=2 \lambda^2 (C_2+C_3-C_1)$, 
$C_j$ is the eigenvalue of the Casimir operator on representation
$\Lambda_j$ and $\langle \phi , \eta \rangle = \sum_{ijk} \phi_{ijk} \chi
^{ijk} $.  Unfortunately, $\langle \phi , \eta \rangle $ vanishes unless
all $C_i=0$. The group invariant three point correlation function is not
zero only for short representations and in this case it computes pairing
$\langle \phi , \eta \rangle$.  This pairing is the analog of the square of
Clebsch-Gordan coefficients $\sum_{ijk}|C^{ijk}_{\mu \nu \lambda}|^2$ for
classical groups.

Similar to the two point correlation functions, one can try to define a
three point correlation function for the individual matrix elements.  We
can compute these correlation functions (i) only for short representations
and (ii) under the assumption that invariant tensors $\phi$ and $\chi$ are
unique (modulo scaling), and in this case they generalize the notion of
Clebsh-Gordan coefficients.  Observe that two and three point correlation
functions turn out to be coupling independent for short representations.

\subsec{Strings on $AdS_3 \times S^3$ and $PSU(1,1|2)$}

In this section we want to make contact with some physical models, in
particular with strings propagating on $AdS$-type backgrounds.  Our
arguments that $PSL(n|n)$ principal chiral model is conformal were purely
algebraic and independent of the choice of the real structure of the
group.  Now, to relate our discussion to strings on $AdS_3 \times S^3$ we
have to consider $PSU(1,1|2)$ principal chiral field.  Similarly, strings
propagating on $AdS_5 \times S^5$ are related to the coset \quo\ of
$PSU(2,2|4)$ principal chiral model.

Let us concentrate on the $PSU(1,1|2)$ case. Our sigma model is non-unitary
and therefore there is no state operator correspondence.  Still both states
and operators can be classified by the representations of the symmetry
group $PSU(1,1|2)_L \times PSU(1,1|2)_R$.  Each individual weight
$\Lambda_L,~\Lambda_R$ is nothing else but a pair of $su(2)$ and $su(1,1)$
weights $\Lambda_L=(\lambda, \mu)_L$ and $\Lambda_R=(\lambda', \mu')_R$
(the weight of $su(2)$ is twice the spin and is always positive integer
$\lambda \in {\bf Z}_+$). To
conform with the physics literature, $\lambda$ denotes the highest weight
of the $su(2)$ while $\mu$ is the {\it lowest} weight, in the sense that it
has the {\it lowest} eigenvalue of $L_0$.

We have to consider both finite and infinite dimensional representations of
the symmetry group. See for example the discussion of
the spectrum of the string theory on $AdS_3 \times S^3$ in
\ref\oog{J. de Boer, H. Ooguri, H. Robins
and J. Tannenhauser, {\it String Theory on $AdS_3$}, hep-th/9812046}.
Not all of the infinite-dimensional representations are highest weight
representations, but we will not have the need for those.
The {\it unitary} infinite-dimensional representations correspond to
normalizable modes in $PSU(1,1|2)$ and non-unitary ones to
non-normalizable. Representation $((\lambda, \mu),(\lambda', \mu'))$ is
unitary if both $\mu$'s are positive. Also note that for a given
representation the sum of the left and right quadratic Casimirs is the
eigenvalue of the Laplacian on the corresponding modes in $PSU(1,1|2)$.

The generic representation $(\lambda, \mu)$ of $psu(1,1|2)$ is irreducible.
But when $\mu = -\lambda$ or $\mu=(\lambda+2)$ the representation turns out
to be reducible 
\foot{ The Kac representation is constructed as a product
$V^{su(2)}_{\lambda} \times V^{su(1,1)}_{\mu} \times {\bf
C}(\theta_1,\theta_2,\theta_3,\theta_4)$, where $V^{su(1,1)}_{\mu}$ is a
finite or infinite dimensional l.w.s module corresponding to weight $\mu$.}.
In this case the irreducible representations are smaller (one can find the
discussion of reducible finite dimensional representations for $psl(2,2)$
in the appendix). In the physical language these are the short
representations, similar to those that appear in the description of BPS
states. For the future let us denote the {\it unitary} short representations
with the following weights as 
$[\lambda, \lambda'] \equiv ((\lambda, \lambda+2)_L, 
(\lambda',\lambda'+2)_R)$.

The sigma model operators $V_{(\Lambda, \Lambda')}$ would enter into the
construction of string vertex operators.  For both $\Lambda$ and $\Lambda'$
short representations their quadratic Casimirs vanish. We can write
it as
\eqn\onsh{(\Delta - \lambda (\lambda +2)-\lambda' (\lambda' +2))
V_{(\Lambda, \Lambda')}=0~,}
where $\Delta$ is the Laplacian operator on $AdS_3$.  This is nothing else
but the supergravity mass shell condition.  In other words, operators
$V_{(\Lambda, \Lambda' )}$ correspond to the massless excitations around
the supergravity background. As it was explained in 
\ref\deboer{J. de Boer,
{\it Six Dimensional Supergravity on $S^3 \times AdS_3$ and 2d Conformal
Field Theory}, hep-th/9806104},
the only representations arising in supergravity have left and right
weights corresponding to the short representations and their $\{su(2)_L,
su(2)_R\}$ weights are 
$\{\lambda,  \lambda \}$, $\{\lambda \pm 1, \lambda \}$ or 
$\{\lambda \pm 2, \lambda \}$, 
where the smaller weight is positive. This restriction
comes from the fact that the supergravity sector contains fields of spin
two at most.  Larger differences would correspond to the higher spin
fields. In particular, the short representations with equal $su(2)$ weights
correspond to the modes of massless scalars in supergravity. 
The weights $((\lambda, \lambda+2),(\lambda, \lambda+2))$ 
labels normalizable modes,
while $((\lambda, -\lambda),(\lambda, -\lambda))$ labels non-normalizable
modes which lie in the highest/lowest weight representations. 
Clearly, the spectrum of the stringy modes is much more complicated.

The conformal field theory on the boundary of $AdS_3$ possesses left and
right ${\cal N}=4$ superconformal algebras. Those have left and right
$PSU(1,1|2)$'s as finite subalgebras. Indeed, one can easily identify them:
$J^3, J^{\pm}$ and $L_0, L_{\pm 1}$ generate $su(2) \times su(1,1)$
subalgebra and $G_{\pm 1/2}^a$ and ${\bar G}_{\pm 1/2}^a$ are the fermionic
generators. The chiral multiplets of the boundary CFT then correspond to
the short representations of $PSU(1,1|2)_L \times PSU(1,1|2)_R$.
The two and three-point correlation functions of chiral operators in the
boundary theory were computed in
\ref\mf{D. Freedman,
S. Mathur, A. Matusis and L. Rastelli, 
{\it Correlation functions in the $CFT_d/AdS_{d+1}$ correspondence},
hep-th/9804058}
\ref\sei{S.~Lee, S.~Minwalla, M.~Rangamani and N.~Seiberg, 
Adv. Theor. Math. Phys. {\bf 2}, 697 (1998), 
hep-th/9806074}
and they are given by the coupling-independent (in the appropriate
normalization) expressions proportional to Clebsch-Gordan coefficients. 
As we found, the two and three-point {\it worldsheet} correlation functions
of the vertex operators corresponding to short representations $(\Lambda,
\Lambda)$ are also independent of the coupling and uniquely determined by
the group structure.

To compute correlation functions of boundary CFT in $AdS$ supergravity, one
has to find the solution of SUGRA equations of motion which approaches
given boundary values at infinity
\ref\withol{E. Witten,
Adv. Theor. Math. Phys. {\bf 2}, 253 (1998), hep-th/9802150}. 
Consider a massless scalar
field in supergravity. For its Kaluza-Klein harmonic lying in the
representation with $su(2)$ weight $\lambda$, the apparent mass in $AdS_3$
is $m^2 = \lambda (\lambda +2)$. For such a scalar the solution approaching
a $\delta(x)$ on the boundary was found in \withol. 
Remarkably,
this solution is exactly the {\it lowest weight state} in the
representation of the $AdS_3$ isometry group $SU(1,1)_L \times SU(1,1)_R$
with both weights equal to $\mu = 1+\sqrt{1+m^2} = \lambda +2 $
\ref\alb{V. Balasubramanian, P. Kraus, A. Lawrence,
Phys.Rev. D59 (1999), hep-th/9805171}. 
Thus it is actually a
state of the lowest $SU(1,1)$ weights in the short representation of the
$PSU(1,1|2)_L \times PSU(1,1|2)_R $ which we earlier denoted $[\lambda,
\lambda]$. Therefore we can identify the vertex operator representing the
same state in $V_{[\lambda, \lambda]}$ as the one which corresponds to the
insertion of the boundary operator at the origin. Let us call this operator
$V_{[\lambda, \lambda]}(z, \bar z|0)$ where $z$ is the world-sheet
coordinate and $0$ is, in a sense, a boundary coordinate. We suppress
the indices labelling the states inside the $V_{[\lambda, \lambda]}$ module.

To shift the above operator from the origin, we can, of course, act on it
with $su(1,1)$ lowering operators $L_{-1}$ and $\bar L_{-1}$ and
thus define
\eqn\tesvar{V_{[\lambda, \lambda]}(z, \bar z|x, \bar x)=\sum {x^n \over
n!}{{\bar x}^m \over m!} [L_{-1},[ \bar
L_{-1}, V_{[\lambda, \lambda]}(z, \bar z)]]~,}
where $(z, \bar z)$ are worldsheet coordinates and $(x, \bar x)$ are now
boundary coordinates. Notice, that one can equally well generalize 
these formulas in the case of different left and right short representations.

Clearly the general string amplitudes are given by complicated expressions
that  involve ghost fields and  worldsheet integrations over positions
of vertex operators, but two- and three- scattering amplitudes are given by
simple minded expressions -- no ghost fields and, thanks to worldsheet
$sl(2,{\bf C})$, no integrations. Therefore we can completely neglect
the worldsheet dependence of the correlation functions assuming that our 
operators are inserted say at $0$  and $1$ for propagator and at  $0$, $1$  
and $\infty$ for the three point scattering.
 
Now we can  write two- and three-point correlation functions for our vertex
operators. We find that the dependence on the boundary coordinates produces exactly
right scaling behavior 
\eqn\corrth{\eqalign{ 
\langle V_{[\lambda, \lambda]}(0|x, \bar x) V_{[\mu, \mu]}( \infty|y, \bar y)  \rangle
&={ \delta_{\lambda \mu} \over |x-y|^{2 (\lambda+2)}}\cr
 \langle V_{[\lambda, \lambda]}(0|x, \bar x) V_{[\mu, \mu]}(1|y, \bar y) 
  V_{[\nu, \nu]}(\infty|u, \bar u) \rangle
&={ T_{\lambda \mu \nu} \over   
|x-y|^{\gamma_{xy}}
|x-u|^{\gamma_{xu}}
|y-u|^{\gamma_{yu}} }~, 
}}
where $\gamma_{xy}= (\lambda+\mu+2 - \nu),~
\gamma_{xu}= (\lambda+ \nu+2 -\mu),
~\gamma_{yu}=(\nu +\mu+2 -\lambda)$
\foot{Tensors $\delta_{\lambda \mu}$ and  $T_{\lambda \mu \nu}$
have suppressed indices that label the states inside representations}.
The coefficients $ T_{\lambda \mu \nu}$ are coupling 
independent (!)
and can be computed in terms of Clebsh-Gordan coefficients. 
Strictly, we can not make computations in the
case of infinite dimensional representation. 
Still, morally speaking this answer is very similar to that 
presented at the end of the previous section. 
Both answers are coupling constant independent
and are expressed in terms of Clebsh-Gordan coefficients
(compare \corrth\ and \sei).

\newsec{Discussion}

We believe that by now we have convinced the reader that  $PSL(n|n)$
sigma model is quite remarkable. 
It gives rise to a conformal field theory. 
This conformal field theory 
is non-trivial, but it contains a certain subsector corresponding to
short representations that is easy to analyze. These are short
(atypical)
representations that correspond to chiral primary fields in the 
boundary theory. The long representations are difficult to analyze and they
give rise to stringy modes (in the corresponding string theory). 
In a sense, the subsector of short representations is  
reminiscent to the ground ring of $c=1$ model. 
The chiral algebra of this sigma model is not just a Virasoro algebra but
its extension similar to $W$ algebras.   
We believe that the study of the representations of this chiral algebra will 
be an important ingredient in the solution of the theory.

It follows from our presentation that all group invariant correlation
functions are coupling constant independent, including the higher
genus calculations. For example, the one loop partition function is
coupling independent and after eliminating the contribution of  
zero modes is equal to $Z'=\eta^2 (q)$ (this is just a contribution of
fermionic $(b,c)$ system). This partition function does not say much about
the spectrum of the theory. It is clear that various twisted version of one
loop partition function would contain more information about the spectrum.
This question is currently under investigation.

Apparently there is another infinite series of supergroups that have a
chance of being conformal -- the $Osp(2n+2|2n)$ groups.  This group
contains $SO(2n+2) \times Sp(2n)$ as a bosonic subgroup.  The dual Coxeter
number vanishes for these groups and therefore the one loop beta function
is zero. We believe that this group has unique rank 3 totally antisymmetric
tensor, which would imply that the theory is exactly conformal. It will be
very interesting to analyze this series.

One can try to go even further. The $PSL(n|n)$ principal chiral model is a
generalization of the $G \times G$ sigma model to a case of a very
special supergroup. It may be promising to generalize $O(N)$
sigma models in a similar way. We only note that the action for such
model looks similar to the action found in \bvw.

As we already mentioned in the introduction, the $PSL(n|n)$ group manifold
is in a sense a Calabi-Yau manifold. One of the directions for future work
could be the construction of other supermanifolds that give rise to 
conformal field theories. For example, various quotients over 
special subgroups also yield the conformal field
theories\ref\bzh{M. Bershadsky and S. Zhukov, {\it in preparation}}. 

There is another observation which is somewhat beyond
the scope of this paper but nevertheless is quite interesting. 
Consider a four dimensional gauge theory based on a supergroup $PSL(n|n)$.
In other words, there is no supersymmetry in four dimensions, but there are
two kinds of vector particles -- the usual bosonic  as well as fermionic. 
It is a four dimensional non-unitary theory. It is possible that certain
version of this theory might appear in the description of the system 
that contains both D-branes and anti D-branes. Now, according to the
arguments presented in this paper, this theory is going to be 
exactly conformal! There is not much difference in group structure
between two dimensional and four dimensional Feynman diagrams.
Therefore, we can repeat all the steps of our proof in the case of this four
dimensional theory. Still, it is unclear whether one can make sense out of
this theory.

It is a challenge to understand the $PSL(n|n)$ sigma models and we hope to
return to this subject in the future.

\bigskip
{\bf Acknowledgments: }
We would like to thank A. Bernstein, S. Coleman, P. Etingoff, D. Kazhdan, 
D. Kutasov, N. Nekrasov, H. Ooguri, V. Serganova, A. Strominger, 
S. Gubser, C. Vafa and J. Zinn-Justin for many valuable discusions. This
research was supported by NSF grant PHY-92-18167.
In addition, the research of M.B is supported in addition by
NSF 1994 NYI award and the DOE 1994 OJI award.

\vfill
\eject

\newsec{Appendix. Representations of $psl(n|n)$}

The simple Lie superalgebra $psl(n|n)$ (or $A(n-1|n-1)$ in Kac's
notation \ref\kac{V.~Kac, Adv. Math. 
{\bf 26}~(1977), 8-96.})
stands out among other superalgebras of the $A(m|n)$
series in many ways and is relatively less studied. Here we will discuss some
of its features.

First of all, $psl(n|n)$  it is not a subalgebra of the matrix superalgebra
$gl(n|n)$ unlike all other algebras of the $A$ series. Indeed, the traceless 
subalgebra $sl(n|n)$ of $gl(n|n)$  is not simple since it has a
non-trivial center $Z={\bf C}\cdot I$ generated by the identity matrix $I
\in sl(n|n)$. The quotient superalgebra $psl(n|n)=sl(n|n)/Z$ is simple for
$n>1$. 

The projection 
\eqn\proj{p: sl(n|n) \to psl(n|n) }
cannot be split (i.e.~$psl(n|n)$ cannot 
be embedded into $sl(n|n)$ as a subalgebra) and therefore $sl(n|n)$ is a
non-trivial central extension of the algebra $psl(n|n)$. This is the only
series of basic classical Lie superalgebras admitting a non-trivial central
extension (the only other examples being the series of ``strange''
superalgebras $Q(n)$ and Hamiltonian superalgebras $H(n)$). 
This phenomenon places $psl(n|n)$ in one class with the Virasoro or current
Lie algebras and shows that not every irreducible representation of the
matrix superalgebra $sl(n|n)$ factors through the projection \proj\
and gives a representation of $psl(n|n)$, but only those with vanishing
``central charge'' (the eigenvalue of $I$).

Another peculiarity of $psl(n|n)$ is that it has rank $2n-2$ (the dimension of
the Cartan subalgebra) and $2n-1$ simple roots (i.e. its simple roots are
linearly dependent).

Before discussing representations of $psl(n|n)$ let us fix some notation. 
Since $psl(n|n)$ does not have a natural matrix representation we will be
working with it in terms of $sl(n|n)$ or $gl(n|n)$ generators keeping in mind
that two matrices from $sl(n|n)$ whose difference is a scalar multiple of $I$
represent the same element of $psl(n|n)$. 
The Cartan subalgebra $H_s$ of $sl(n|n)$ has dimension $2n-1$ and is spanned
by elements 
\eqn\cartan{\eqalign{h_i &= E_{ii}-E_{i+1,i+1}, \quad 1\leq i \leq n-1, \ 
        {\rm and} \ n+1\leq i \leq 2n-1~, \ {\rm and} \cr
h_n &= E_{nn}+E_{n+1,n+1},
}}
where $E_{ij}$ is the elementary matrix whose only non-zero entry is $1$ on
the intersection of the $i$-th row and the $j$-th column. Elements $E_{ii} \quad
1\leq i \leq 2n $ generate the Cartan subalgebra $H_g$ of $gl(n|n)$. 

The remaining part of the root decomposition is the same for both
$sl(n|n)$ and $gl(n|n)$ and is given 
by the following root vectors and the corresponding roots:
\eqn\roots{\eqalign{
E_{ij}~, \quad \quad \qquad \epsilon_i - \epsilon_j, \cr
E_{n+i,n+j}~, \quad \qquad \quad \delta_i - \delta_j, \cr
E_{i,m+j}~, \quad \qquad \quad \epsilon_i - \delta_j, \cr
E_{m+i,j}~, \quad \qquad \quad \delta_i - \epsilon_j,
}}
$i,j=1,\ldots,n$, where $\epsilon_i, \delta_i \in H_g^*$ are functionals on $H_g$ 
such that $\epsilon_i(x)= x_i$ and $\delta_i(x)=x_{n+i}$ for
$x={\rm diag}(x_1,\ldots,x_{2n}) \in H_g$. 
The roots $\epsilon_i - \delta_j$ and $ \delta_i - \epsilon_j$ are odd (which
means that the corresponding root vector is odd), the remaining
ones are even. The so-called distinguished system of simple roots is
chosen as follows:
\eqn\simple{\eqalign{  
\alpha_i&=\epsilon_i-\epsilon_{i+1}, \quad 1 \le i \le n-1~, \cr
\alpha_n &= \epsilon_n-\delta_1, \cr
\alpha_{n+i}&=\delta_i-\delta_{i+1}, \quad 1 \le i \le n-1~.
}}

With this choice of simple roots, the $2n^2-n$ positive roots
of $sl(n|n)$ are $E_{ij}$ for $i<j$, $E_{n+i,n+j}$ for $i<j$, and  
$E_{i,m+j}$. In the distinguished system there is  only one odd simple root
$\alpha_n$.

There are two different ways to represent $sl(n|n)$ weights (i.e.~,elements of
$H_s^*$) in coordinates both of which have some advantages. 
First, we can express $\Lambda \in H_s^*$ in terms of the basis \simple\
\eqn\acoord
{\Lambda = [a_1,a_2,\ldots,a_{2n-1}]=\sum_{i=1}^{2n-1} a_i \alpha_i, \quad
{\rm where} \ a_i = \Lambda(h_i)~.}
On the other hand, since $H_s^*$ is a quotient of $H^*_g$ by the element 
$$
\omega_0=\sum_{i=1}^{n}\epsilon_i - \sum_{i=1}^{n}\delta_i~,
$$
we can represent $\Lambda$ in terms of $\epsilon_i$ and $\delta_i$ as
\eqn\lamu
{\Lambda = (\lambda_1,\ldots,\lambda_n;\mu_1,\ldots,\mu_n)=
\sum_{i=1}^{n} \lambda_i \epsilon_i + \sum_{i=1}^{n} \mu_i
\delta_{i}~,}
keeping in mind that the strings
$\Lambda=(\lambda_1,\ldots,\lambda_n;\mu_1,\ldots,\mu_n)$ and
$\Lambda + \mu = (\lambda_1+1,\ldots,\lambda_n+1;\mu_1-1,\ldots,\mu_n-1)$
represent the same element in $H^*_s$.
The relation between the two coordinate systems is given by
\eqn\bases{(\lambda_1,\ldots,\lambda_n;\mu_1,\ldots,\mu_n) = 
[\lambda_1-\lambda_2,\ldots,\lambda_{n-1}-\lambda_n,
\lambda_n+\mu_1,\mu_1-\mu_2,\ldots,\mu_{n-1}-\mu_n].}
The dual space $H^*$ of the Cartan subalgebra $H$ of $psl(n|n)$ is a
codimension one subspace of $H_s^*$ that consists of linear functionals on
$H_s$ vanishing on the vector $I \in sl(n|n)$.  
The equation
$$
I = h_1 + 2h_2 + 3h_3 + \ldots nh_n - (n-1)h_{n-1}- \ldots -2h_{2n-2}-h_{2n-1},
$$
shows that 
$[a_1,a_2,\ldots,a_{2n-1}]$ belongs to $H^*$ if and only if 
\eqn\constr{
\alpha_1+2\alpha_2+\ldots+n\alpha_n+(n-1)\alpha_{n-1}+\ldots + 2\alpha_{2n-2}
+ \alpha_{2n-1} = 0~.
}
and, therefore,
$$
a_n = -{1 \over n}\big(\sum_{i=1}^{n-1} i a_i - \sum_{i=1}^{n-1} (n-i)a_{n+i}\big)~.
$$
In the $(\lambda,\mu)$ form, the equation \constr\ looks simpler
$$
\sum_i (\lambda_i + \mu_i) = 0.
$$

The equation \constr\ implies in particular
that the simple roots \simple\ of $psl(n|n)$ are not linearly independent.
\medskip

Since $psl(n|n)$ is a quotient of $sl(n|n)$, every representation of 
$psl(n|n)$ is automatically a representation of $sl(n|n)$. The relationship
between representations of $sl(n|n)$ and $gl(n|n)$ is slightly different
because $sl(n|n)$ is only a subalgebra and not a quotient of $gl(n|n)$
(which is the case with $gl(m|n)$ for $m\ne n$). Therefore, every
representation of $gl(n|n)$ is a representation of $sl(n|n)$, but the
converse is not true in general. However, since every finite-dimensional
irreducible representation of $sl(n|n)$ is a highest weight representation
$V_\Lambda$, it is determined by a  weight $\Lambda \in H^*_s$ (which can be
considered as a one-dimensional
representation of the Cartan subalgebra $H_s$). But every
weight $\Lambda \in H_s^*$ can be extended (not uniquely) to  a weight
$\tilde \Lambda \in H^*_g$ and thus the action of $sl(n|n)$ on $V_\Lambda$
can be extended to a $gl(n|n)$-action.  This allows us to work with
irreducible representations of $psl(n|n)$ in terms of $sl(n|n)$ or $gl(n|n)$
representations. In particular, every irreducible representation of
$psl(n|n)$ lifts to an irreducible representation of $gl(n|n)$. 

Let us recall Kac's construction of representations of $sl(n|n)$
(which works with some modifications also for other  basic classical Lie
superalgebras).
Denote $L=sl(n|n)$ and let $L= L_{-1} \oplus L_0 \oplus L_1$, where 
\eqn\even
{L_0=sl(n) \oplus {\bf C} \oplus sl(n) }
is the even subalgebra of $L$ and $L_{\pm 1}$ the subalgerbras
corresponding to upper- and lower-triangular odd matrices in $sl(n|n)$.
Pick a representation of the subalgebra $L_0$ and extend it to a representation
of the subalgebra  $K=L_0\oplus L_1$ by setting $L_1 V = 0$. 
The {\it Kac representation }
\eqn\kac{%
\tilde V = {\rm Ind}^L_K V = {\cal U}(L) \otimes_{{\cal U}(K)} V
}
corresponding to $V$ (also called the Kac or 
induced module) as a vector space is isomorphic to the tensor product
$$
V \otimes \bigwedge L_{-1},
$$
where $\bigwedge L_{-1}$ is the Grassman algebra of the vector space $L_{-1}$.

If we start with an irreducible representation $V=V_\Lambda$
of the even subalgebra \even\  corresponding to a 
a weight vector 
\eqn\weight
{\Lambda=[a_1,a_2,\ldots,a_{n-1}, c,
b_1,b_2,\ldots,b_{n-1}]  \in H^*_s~,
}
where $[a_1,a_2,\ldots,a_{n-1}]$ and $[b_1,b_2,\ldots,b_{n-1}]$
are dominant $sl(n)$ weights (i.e. all $a_i$ and $b_j$ are non-negative
integers), then the Kac module 
$ V(\Lambda) = \tilde{V}_\Lambda$ 
has dimension 
$$
2^{n^2}\prod_{1\leq i \leq j \leq n-1}
{ {\big((a_i+1)+\ldots+(a_j+1)\big) \big( 
(b_i+1)+\ldots+(b_j+1)\big)} \over (j-i+1)^2  }.
$$

For a generic ({\it typical}) weight  $\Lambda$ the representation 
$V(\Lambda)$ is irreducible, otherwise it has a unique maximal invariant
subspace $U$ and the quotient representation $M(\Lambda)=V(\Lambda)/U$ is
irreducible. 
All irreducible representations of $sl(n|n)$ can be obtained by
this construction. 
If the Kac representation
\kac\ is reducible, the weight
$\Lambda$ is called {\it atypical}. This happens exactly when
\eqn\typical{
a_i+a_{i+1}+\ldots+a_n-a_{n+1}-\ldots -a_{n+j-1} + i-j-n+1 = 0
}
for some $1\leq i,j \leq n$.
This condition means that at least one of the inner products
$$
\langle \Lambda + \rho | \sigma_{ij}\rangle
$$
vanishes, where $\sigma_{ij}=\epsilon_i-\delta_j$ a positive odd root and 
\eqn\rro
{\rho=[1,1,\ldots,1,0,1,\ldots,1]={1 \over 2}(2n-1,2n-3,\ldots,-2n+1)} 
is the half-sum of the positive roots.

In terms of the $gl(n|n)$ coordinates \lamu\ the condition \typical\ can be
rewritten as
\eqn\typlm
{\lambda_i+\mu_j+n-i-j+1=0~.}
For example, the Kac representation $V(\Lambda)$ of $sl(2|2)$ with 
$\Lambda=[a,c,b], \quad a,b \in {\bf Z}_+$ is atypical if $c$ is equal to one
of the following 
\eqn\slatyp{0, \ -a-1, \ b+1, \ b-a.}
Now when does a highest weight 
representation of $sl(n|n)$  descend to a representation
of $psl(n|n)$ via projection \proj~? This happens when the weight
$\Lambda$ belongs to the subspace of $H^*$ given by the constraint
\constr. This gives the following equation for the component $c$ of the
weight \weight\
\eqn\ppi{
c = - {1 \over n}\big(\sum_{k=1}^{n-1} k a_k  - \sum_{k=1}^{n-1} k
b_{n-k+1}\big)~.  
}
Therefore, irreducible representations of $psl(n|n)$ are determined by a pair
of $sl(n)$ weights $\alpha=[a_1,\ldots,a_{n-1}]$ and
$\beta=[b_1,\ldots,b_{n-1}]$. 

This shows another feature of $psl(n|n)$ that distinguishes it from other
algebras of the $A$ type (and from most of other simple Lie superalgebras)
--- its irreducible representations depend on $2n-2$ integer parameters and
do not have continuous parameters. (Irreducible representations of $sl(m|n)$
with $m \neq n$ 
can be deformed, because there are no restrictions on the coordinate $c$ of
the highest weight.)

In the $psl(2|2)$ case for example, 
we have from \ppi\ that $c={(b-a) \over 2}$ and the atypicality condition
\slatyp\ now becomes just $a = b$. The Kac module $V_{\Lambda}$ is reducible
when $\Lambda=[a,0,a]$. Let us denote the corresponding
 irreducible representation by $M_a$. 
If $a \neq 0,1$ the structure of the module $V_\Lambda$ is very simple,
namely
\eqn\emb{ M_a \subset K_a \subset V_{\Lambda}~,}
where $ V_{\Lambda}/ K_a= M_a$ and 
$K_a/ M_a= M_{a+1} \oplus M_{a-1}$.

For $psl(3|3)$ there are six  
different types of atypicality some of which can happen simultaneously:
\eqn\psltt{\eqalign{
a_1+2a_2-2a_4-a_5&=0, \cr
-a_1+a_2-a_4+a_5&=0, \cr
2a_1+a_2+2a_4+a_5+2&=0, \cr
2a_1+a_2-a_4+a_5+1&=0, \cr
-a_1+a_2+2a_4+a_5+1&=0, \cr
-a_1+a_2-a_4-2a_5-1&=0.
}}

\newsec{Appendix. Casimir operators of $psl(n|n)$ }

The algebra of casimirs --- the center of the universal enveloping algebra
--- for any Lie superalgebra $L$ with an invariant inner product is isomorphic 
to the algebra of invariant polynomial functions on $L$.
For classical matrix Lie superalgebras this algebra $I(L)$ of invariants
is well studied (see, e.g. \ref\sergeev{A.~Sergeev, 
C.R.~Acad. Bulgare Sci. {\bf 35} (1982), 573; 
{\it The invariant polynomials on simple Lie superalgebras}, 
math.rt/9810111}). 
An analog of the classical
Chevalley's theorem describes $I(L)$ in terms of restrictions of the
invariant polynomials to the Cartan subalgebra $H$ of $L$.
For $L=sl(n|n)$ for example, $I(L)$ is generated by polynomials $t_n$ where
$t_k(X) = str (X^k)$ for $X \in 
sl(n|n)$ \quad $k=2,3,\ldots$~. Only  the first $2n-1$ functions $t_k$ are
algebraically independent, the rest being {\it rational} functions of the
first $2n-1$ ones. However, when $n \to \infty$, all $t_k$ become
algebraically independent.

The case of $psl(n|n)$ is more interesting and difficult.
Since every invariant polynomial on $psl(n|n)$ is also an invariant
polynomial on $sl(n|n)$, the algebra of Casimirs for $psl(n|n)$ is a
subalgebra $A$ of $B={\bf C}[t_1,t_2,\ldots,t_k,\ldots]$ that consists 
of all polynomials in $t_1,t_2,\ldots$ that can be pushed to a well-defined
function on $psl(n|n)$, i.e. 
\eqn\casalg{
  A=\{ f \in B | f(X+kI)=f(X), \quad ~{\rm for\ any\ } X \in sl(n|n)\ , k \in
  {\bf C}\}~. 
}
This is a rather strong condition and
a priori  it is not even clear whether any non-constant function with this
properties should exist. 

It turns out, however, that the algebra $A$ of Casimirs for $psl(n|n)$ is
quite large and has a rich and interesting structure.

First, it is easy to check that the quadratic polynomial $t_2$ belongs to $A$. 
It corresponds to the invariant inner product on $psl(n|n)$. 

To construct new Casimirs of higher order let us consider the following 
formal power series in infinitely many variables $s_1,s_2,\ldots$:
\eqn\series{
f(X;s_1,s_2,\ldots) = {\rm str}(e^{s_1 X}) {\rm str}(e^{s_2 X})\ldots . 
}
If only finitely many, say the first $p$, 
of the variables $s_1,s_2,\ldots $ are non-zero and $\sum_{k=1}^p s_k =0~,$
then, obviously,
 $$
f(X;s_1,s_2,\ldots)=f(X+kI;s_1,s_2,\ldots).$$
 Therefore,
the coefficient $c(a_1,a_2,\ldots,a_{p-1})$ of $f$ at $s_1^{a_1}
s_2^{a_2} \ldots s_{p-1}^{a_{p-1}}$ after expanding $f$ in $s_1,
s_2,\ldots,s_{p-1}$ belongs to the algebra of Casimirs $A$.

For example,
\eqn\second{
c(2p) = 
2 d_{2}d_{2p-2} - 2 d_{3}d_{2p-3} + \ldots + (-1)^p d_p^2~, 
}
where we set $d_k = t_k/k!$ so that $str(e^{s_i X}) = \sum_{k \ge 2}s_i^k
d_k$, since $t_0=str I = 0$ and $t_1=str X =0$.
Analogously we have invariants with more terms in the products, for example
$$
c(p,2q)=
\sum_{0 \le i \le p} \sum_{0 \le j \le 2q} (-1)^{p+2q-i-j} C^{p+2q-i-j}_{p-i} 
d_i d_j d_{p+2q-i-j}
$$ 
and, in particular, 
$$
c(3,6) = -3d_2^2d_5+3d_2d_3d_4-d^3_3~.
$$
Here is the simplest example of an invariant involving products of 
four $t_i$'s that does not belong to the subalgebra generated by
$c(2p)$ and $c(p,2q)$:
\eqn\fourth{\eqalign{
c(3,4,8)&= -252d_2^3 d_9+ 252d_2^2d_3d_8 - 98d_2^2 d_4 d_7+ 30 d_2^2 d_5 d_6
- 77d_2 d_3^2 d_7 \cr
 &+68 d_2 d_3 d_4 d_6 -
30 d_2 d_3 d_5^2 - d_2 d_4^2 d_5 + 3 d_3^3 d_6 - 3 d_3^2 d_4 d_5 + d_3 d_4^3~. 
}}

It can be proved that  the whole algebra of Casimirs is generated
by $t_2$ and the invariants
\eqn\casim
{c(a_1,a_2,\ldots,a_{m-1},2a_m) \quad {\rm with} \ 
3 \leq a_1 < a_2 < \ldots < a_{m-1} \leq a_m~.
}
The leading in the lexicographical order of this invariant is
$d_{a_1}d_{a_2}\ldots d_{a_{m-1}}d_{a_m}^2 $.

For $psl(2|2)$ we have $t_{2k+1}=0$, and therefore, all invariant polynomials
have even degrees.
In this case, the invariants $t_2$ and $c(6)$ are algebraically independent,
while all the other are rational functions of these two.
For example,
\eqn\highcas{\eqalign{
t_2 ^2 c(8)&={2 \over 5}c(6)^2 +{1 \over 720}t_2^6~, \cr
 t_2^4 c(10)&={3 \over 35} c(6)^3 +{1 \over 1008} c(6)t_2^6~.
}}
For $n \ge 2$  only $2n-2$ of the invariants \casim\ are algebraically
independent, but for $n \to \infty$ they become algebraically independent,
and $A$ a free algebra with infinitely many generators.

\listrefs

\end